\documentstyle[epsfig]{aipproc}

\newcommand{\etal}{{\it et~al.\/}}
\newcommand{\Hline}[1]{\mbox{H{\footnotesize {#1}}}}
\newcommand{\Halpha}{\Hline{\mbox{$\alpha$}}}
\newcommand{\HII}{\mbox {H\thinspace{\footnotesize II}}}

\begin{document}

\title{The Starburst Intensity Limit And Its Ultraviolet Implications}

\author{Gerhardt R.\ Meurer}

\address{The Johns Hopkins University \\
Department of Physics and Astronomy \\
Baltimore, MD 21218}

\maketitle

\begin{abstract}
Our recent work on starbursts, particularly in the ultraviolet (UV), is
summarized.  The intrinsic UV fluxes of UV selected starbursts can be
derived from UV data alone because, to first order, their dust behaves
like a foreground screen.  This allows a comparison of the bolometric
effective surface brightness $S_e$ of UV selected starbursts to other
starburst samples.  Starbursts have a robust (90th percentile) upper
limit $S_e \lesssim 2.0 \times 10^{11}\, L_\odot\, {\rm kpc^{-2}}$,
which strongly suggest that their global star formation intensities
are regulated.  The mechanism(s) involved in the regulation are not
yet clear.  The dust attenuation corrections for high-$z$ starbursts
are significant.  Calculations of the rate of evolution in the early
universe based on zero dust interpretations are probably
underestimated by about an order of magnitude.  Hence the early
universe was not quiescent, but obscured.
\end{abstract}


\section*{Introduction}

The pace of cosmic evolution is marked out by the condensation of
stars from the ISM.  Starbursts, regions of intense massive star
formation which can dominate the bolometric output of galaxies, are
crucial to our understanding of this evolution at all redshifts.  In
the local universe about 25\%\ of high mass star formation is
contained within starburst galaxies (Heckman, 1997; Gallego \etal\
1995).  At high redshift ($z > 2$), starbursts are the the easiest
galaxies to detect, and can be used to directly trace the rate of
evolution (Madau \etal\ 1996). So, to understand the cosmological pace
of evolution, we must understand starbursts.  We would like to know
how to derive intrinsic star formation rates and the physics that
determines these rates.  Can the starburst ``knob'' be cranked up to
arbitrary levels or is its amplitude limited?

Recently we imaged a small sample of starbursts in the
ultraviolet (UV) using the HST (Meurer \etal\ 1995; hereafter M95).
The basic UV structure of a starburst consists of prominent compact
star clusters embedded in a more diffuse cloud.  In the nearest
starbursts the diffuse light starts to resolve into stars, hence star
formation within starbursts occurs in both clustered and diffuse
modes. In M95 we compared the effective surface brightness, $S_e$ (the
surface brightness within the radius containing half the total
emission), of starbursts and concluded that there is an upper limit to
$S_e$.  This suggests that the intensity of starbursts is limited.
Lehnert and Heckman (1996) found a similar surface brightness limit in
their combined FIR/\Halpha\ study.  This prompted our more
comprehensive investigation of starburst surface brightnesses (Meurer
\etal\ 1997; hereafter M97) which we report on here.

\section*{Attenuation and Reddening Due to Dust}

Vacuum-UV imaging is indispensable for understanding starbursts since
it allows the direct detection of the hot-high mass stars that power
starbursts.  In contrast other tracers of high-mass stars,
such as \Halpha\ (and other emission lines), radio continuum, and
far-infrared (FIR) emission originate in the ISM and
only indirectly trace star formation.  Dust presents the biggest
obstacle to understanding UV observations, since dust extinction and
scattering are strongest in the UV resulting in a net attenuation of
the UV flux.  Fortunately, our work has proven that UV observations
can reveal essential properties of high mass star formation, even in
the presence of modest amounts of dust.

\begin{figure}[t]
\centerline{\hbox{\epsfig{figure=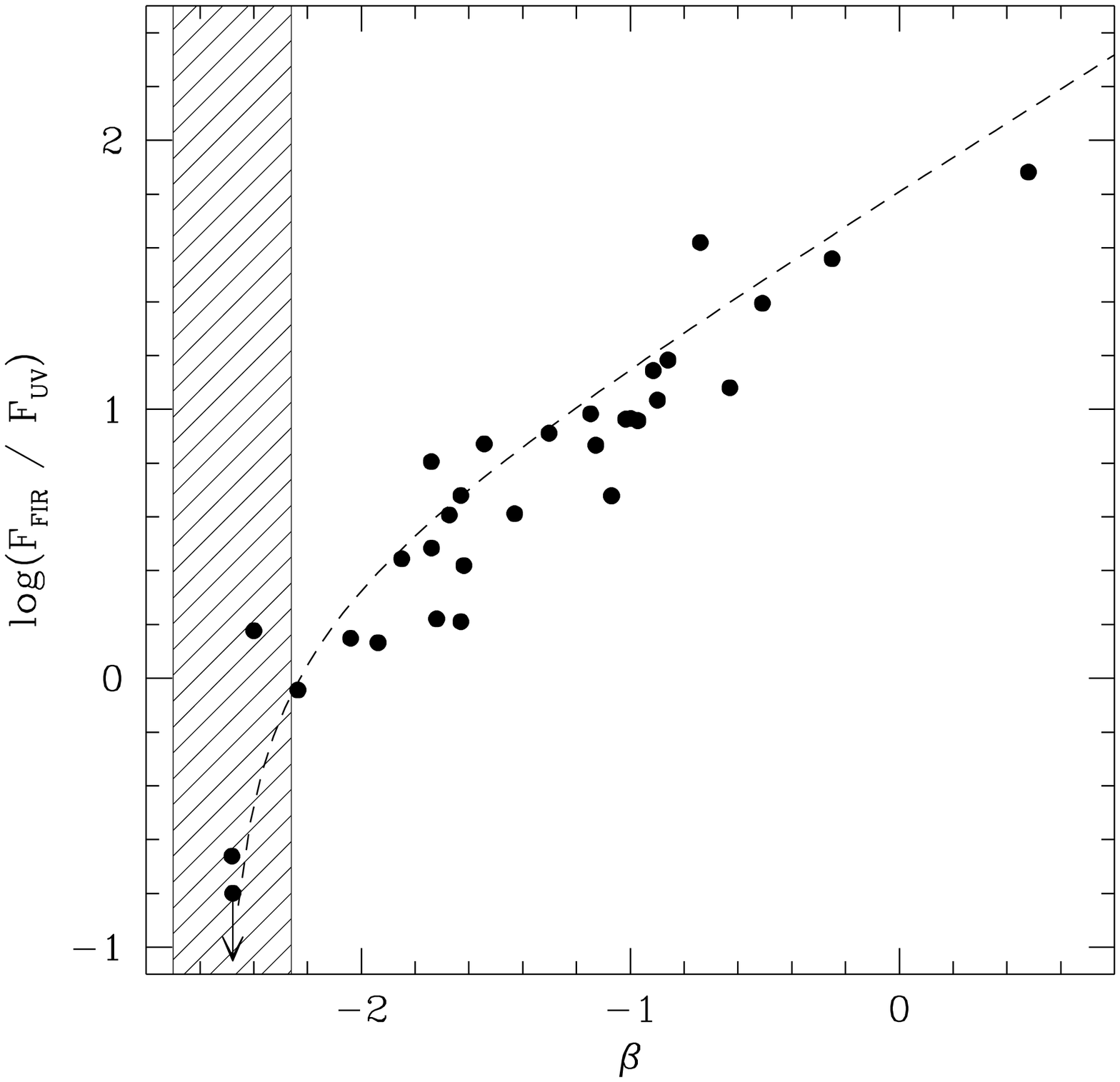,height=7cm}}
\parbox[b]{7cm}{\footnotesize {\bf Figure 1.} The ratio of far
infrared to UV fluxes compared to UV spectral slope $\beta$ (flux
density $f_\lambda \propto \lambda^\beta$) which is equivalent to a UV
color.  The data correspond to UV-selected starbursts.  The fluxes are
measured in in the FIR by IRAS and in the UV ($\lambda \approx
2300$\AA) by HST or IUE.  The hatched region shows the expected
$\beta$ for naked ionizing populations.  The dashed line shows the
expected relationship for a starburst having an intrinsic $\beta_0 =
-2.5$ behind a foreground screen of dust obeying the Calzetti \etal\
(1994) attenuation law. \vspace*{9mm}}}
\end{figure}

We are saved because dust is not a sink for photons; the absorbed
radiation is reemitted thermally in the FIR.  Hence the ratio of FIR
to UV fluxes is an indication of net attenuation.  Figure~1 shows a
positive correlation between UV spectral slope $\beta$ and
$\log(F_{\rm FIR}/F_{\rm UV})$ for a sample of strongly star forming
galaxies as indicated by their strong recombination emission line
spectra.  {\em A priori\/} we expect their UV spectra to be that of an
ionizing population, and hence to have an intrinsic color in the
hatched region of the plot.  This correlation is readily understood in
terms of a simple foreground screen dust geometry, as shown by the
model line.  Although the dust distribution, in exact detail, may not
be a {\em homogeneous\/} foreground screen, and other quantities may
be involved in the correlation, this is largely unimportant for
determining the attenuation correction.  So long as the FIR emission
results from dust reradiating the light absorbed in the UV and
optical, any model that reproduces the correlation in Fig.~1 will
recover the UV flux of ionizing populations to within a factor of
$\sim$3 (corresponding to the spread in $\log(F_{\rm FIR}/F_{\rm
UV})$).

\section*{The Starburst Intensity Limit}

\begin{figure}[t]
\centerline{\hbox{\epsfig{figure=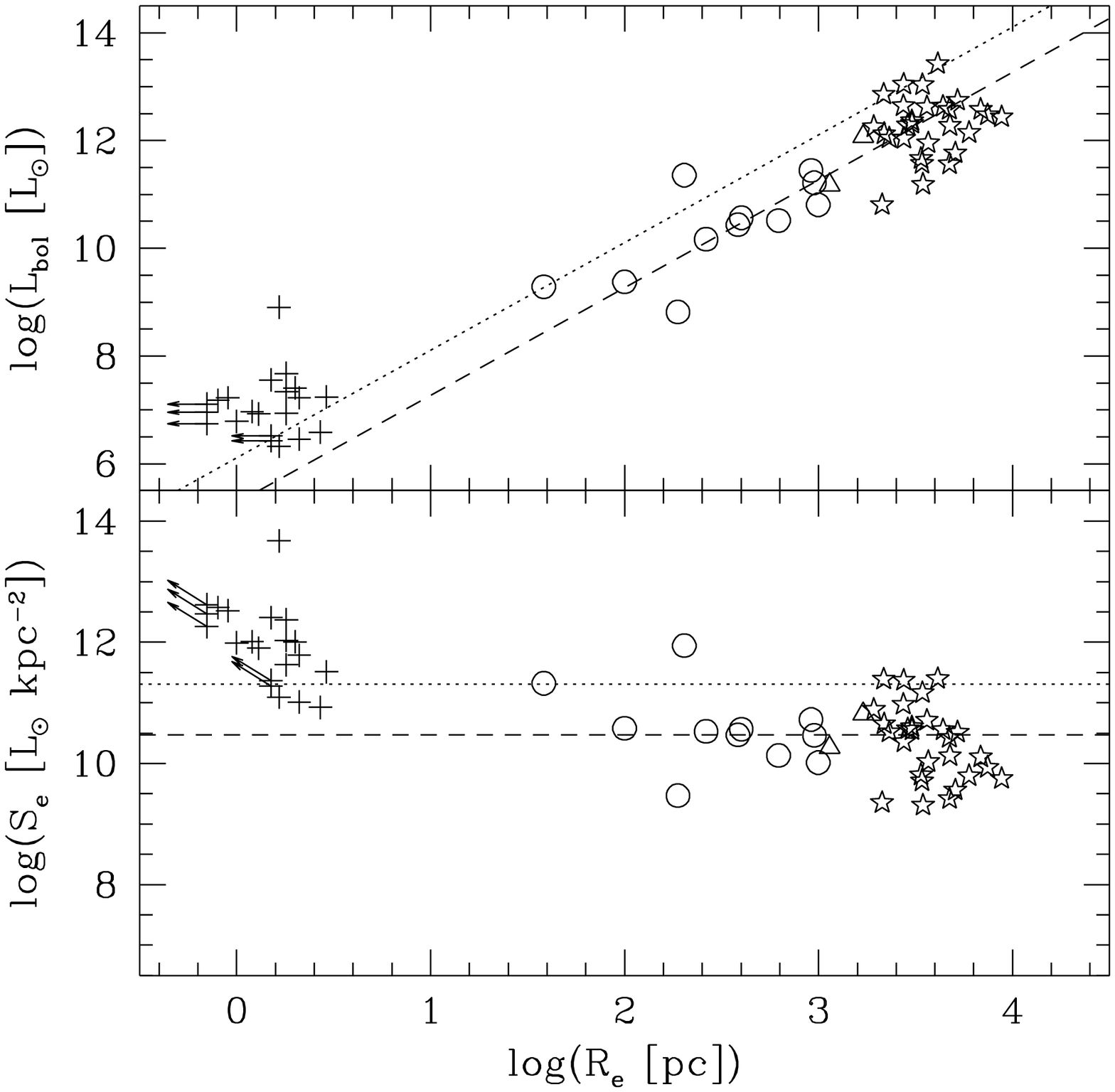,width=7cm}
\epsfig{figure=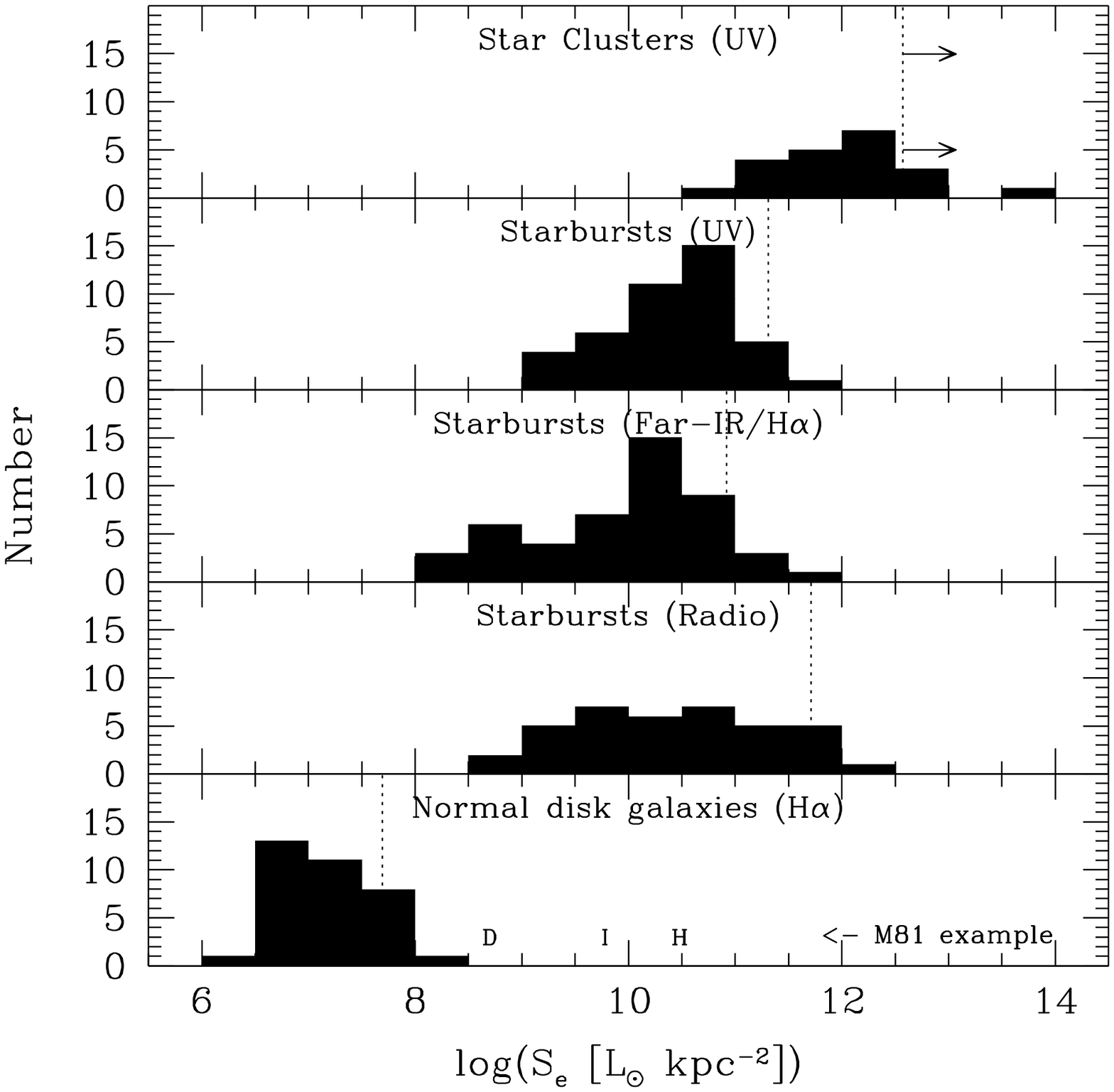,width=7cm}}}
\parbox{140mm}{\footnotesize {\bf Figure 2} (left). Bolometric
luminosity $L_{\rm bol}$ and effective surface brightness $S_e$
plotted against effective radius $R_e$ for UV-selected starbursts and
star clusters.  The dotted and dashed lines are the 90th and 50th
$S_e$ percentiles for the starbursts.  The circles corresponds to
local starbursts ($D < 75$ Mpc), triangles to starbursts at $z = 0.4$,
stars to high redshift ($2.2 < z < 3.5$) starbursts, and crosses
to clusters within starbursts having $D < 10$ Mpc.  \\ 
{\bf Figure 3} (right).  $S_e$ distributions for star clusters (from
Fig.~2), UV selected starbursts (from Fig.~2), FIR selected starbursts
measured in the FIR and \Halpha, FIR selected starbursts observed at
21cm, and normal disk galaxies observed at \Halpha\ (see M97 for
sample details). The dotted lines show the 90th percentiles of the
distributions. In the bottom panel we show preliminary $S_e$
measurements of an UV ($\lambda \approx 1500$\AA) image of M81 at
obtained with the UIT (Hill \etal\ 1992). }
\end{figure}

The evidence for a starburst intensity limit is summarized in Figs 2
and 3.  In Fig.~2 we show measurements of our rest frame UV sample
after dust attenuation correction as outlined above.  For comparison
with the other samples, shown in Fig.~3, luminosities and surface
brightnesses are converted to bolometric units following the models of
Leitherer \&\ Heckman (1995).  Note that the shape of the low $S_e$
end of the distributions in Fig.~3 are determined by (largely
arbitrary) selection effects.  However there is no selection against
high surface brightness.  The three starburst $S_e$ distributions have
90th percentile upper limits within a factor of three of $S_{\rm lim}
\approx 2 \times 10^{11}\, L_\odot\, {\rm kpc^{-2}}$, which we call
the starburst intensity limit.  This limit is very robust, applying to
starbursts with $R_e \sim 0.1$ to 10 kpc, to both local starbursts,
and systems out to redshift $z \sim 3$, to both UV/optically selected
starbursts and FIR selected starbursts, and to observations obtained
in the UV, \Halpha, FIR, and radio continuum.  The lack of a strong
$\lambda$ effect, indicates that it corresponds to a limit on star
formation intensity rather than being an opacity effect, and is
further vindication of the UV attenuation estimates. This limit
strongly indicates that the global intensity of star formation is
regulated within starbursts.

In M97 we explored two physical mechanisms that may be responsible for
the intensity limit: galactic winds (Heckman \etal\ 1990), and
gravitational stability of the inner disk (following Kennicutt, 1989).
Although we find evidence that the intensity limit is related to both
mechanisms, neither predicts the value of the intensity limit, nor its
robustness. Hence {\em determining the physics of the starburst
intensity remains a major theoretical challenge\/}.

Although robust, this limit does not hold for all
star forming scales.  In particular it does not hold
for the star clusters embedded within starbursts.  These are small
$R_e \lesssim 10$ pc (M95) and can have $S_e$ two orders of magnitude
or more intense than $S_{\rm lim}$.  This is not a contradiction
because $S_{\rm lim}$ refers to a global quantity dominated by diffuse
light, whereas individual clusters usually do not dominate the total
luminosity of starbursts.

Star formation in normal disk galaxies is typically three orders of
magnitude less intense than in starbursts, as seen in Fig.~3.
However, this may largely reflect the different star formation
patterns.  Typically star formation fills the central regions of
starburst galaxies, whereas in normal galaxies it usually consists of
\HII\ regions, often arranged in spiral arms or rings, but otherwise
fairly isolated.  We illustrate how large a difference covering factor
can make in the bottom panel of Fig.~3 using a UIT UV image of M81
(Hill \etal\ 1992).  Its disk averaged $S_e$, marked ``D'', is the
surface brightness within the elliptical aperture containing half the
UV light (the same technique used to measure the $S_e({\rm H}\alpha)$
values shown in the bottom panel).  The {\em isophotal\/} $S_e$,
marked ``I'', is the surface brightness within the isophote containing
half the total flux (in other words, pixels having this or higher
surface brightness comprise half the total UV flux). In effect, this
value does not include any of the ``empty space'' between \HII\
regions. The $S_e$ of one of M81's brightest \HII\ regions is marked
``H''.  This comparison suggests that star formation in normal
galaxies, averaged over only the actively star forming regions, may
occur with the same intensity distribution as starbursts.  More
measurements of normal galaxies are needed to test this hypothesis.

\section*{Implications}

Our results have far reaching implications. We demonstrated that
intrinsic UV fluxes of starbursts can be recovered from UV data alone
(Fig.~1).  This confirms the importance of UV observations to the
study of extragalactic high mass star formation.  The dust attenuation
corrections are significant in the UV.  The median $\beta = -1.1$ of
our UV sample corresponds to an attenuation factor of $\sim$ 8 at
$\lambda \approx 2300$\AA\ and $\sim 15$ at $\lambda = 1600$ \AA.  The
latter value corresponds to the rest $\lambda$ of the Lyman drop out
galaxies observed at high redshift (Madau \etal\ 1996).  These have
properties (including $\beta$) consistent with those of local UV
selected starbursts.  Madau \etal\ (1996) use the UV luminosity
density of these systems to infer that the universe was fairly
quiescent (low star formation rate density, or equivalently metal
production rate) at redshift $z \approx 3$.  However they applied no
dust extinction corrections.  After applying our UV attenuation
corrections, and adopting the same cosmology as Madau \etal\ ($H_0 =
50\, {\rm km\, s^{-1}\, Mpc^{-1}}$, $q_0 = 0.5$) we find a lower limit
of $\sim 3 \times 10^{-3} \, {\rm M_\odot\, yr^{-1}\, Mpc^{-1}}$ for
the metal production rate at $z \approx 3$.  This is about six
times higher than their estimate of the Hubble time averaged metal
production rate.  We find that rather than being quiescent we are
observing an active but moderately obscured early universe.

{\footnotesize {\bf Acknowledgements:} I thank my collaborators for
their work in the M95 and M97 papers: Don Garnett, Anne Kinney, Matt
Lehnert, Claus Leitherer, James Lowenthal, Carmelle Robert, and
especially Tim Heckman.  Those papers and this benefited from many
stimulating discussions with Daniela Calzetti.}


\begin{references}

\bibitem{} Calzetti, D., Kinney, A.L., \&\ Storchi-Bergmann, T. 1994,
ApJ, 429, 582 (C94)
\bibitem{} Gallego, J., Zamorano, J., Arag\'on-Salamanca, A., \&\
Rego, M.  1995, ApJ, 455, L1
\bibitem{} Heckman, T.M., Armus, L., \&\ Miley, G.K. 1990, ApJS, 74, 833
\bibitem{} Heckman, T.M. 1997, in Star Formation Near and Far, edited by
S.S.\ Holt \&\ L.G. Mundy, (AIP, Woodbury NY), p.\ 271
\bibitem{} Hill, J.K., \etal\ 1992, ApJL, 395, L37
\bibitem{} Kennicutt, R.C. 1989, ApJ, 344, 685
\bibitem{} Lehnert, M., \&\ Heckman, T.M. 1996, ApJ, 472, 546
\bibitem{} Leitherer, C., \&\ Heckman, T.M. 1995, ApJS, 96, 9
\bibitem{} Madau, P., Ferguson, H.C., Dickinson, M.E., Giavalisco,
M., Steidel, C.C., \&\ Fruchter, A. 1996, MNRAS, 283, 1388
\bibitem{} Meurer, G.R., Heckman, T.M., Leitherer, C., Kinney, A.,
Robert, C., \&\ Garnett D.R. 1995, AJ, 110, 2665 (M95)
\bibitem{} Meurer, G.R., Heckman, T.M., Lehnert, M.D., Leitherer,
C., \&\ Lowenthal, J.  1997, AJ, {\em in press} (M97; astro-ph/9704077)

\end{references}
\end{document}